# Satellite-to-ground quantum communication using a 50-kg-class micro-satellite

Hideki Takenaka, Alberto Carrasco-Casado, Mikio Fujiwara,

Mitsuo Kitamura, Masahide Sasaki*, and Morio Toyoshima

National Institute of Information and Communications Technology (NICT),

4-2-1 Nukui-kitamachi, Koganei, Tokyo 184-8795, Japan.

*e-mail: psasaki@nict.go.jp

**Recent rapid growth in the number of satellite-constellation programs for remote sensing and communications, thanks to the availability of small-size and low-cost satellites, provides impetus for high capacity laser communication (lasercom) in space. Quantum communication can enhance the overall performance of lasercom, and also enables intrinsically hack-proof secure communication known as Quantum Key Distribution (QKD). Here, we report a quantum communication experiment between a micro-satellite (48 kg and 50 cm cube) in a low earth orbit and a ground station with single-photon counters. Non-orthogonal polarization states were transmitted from the satellite at a 10-MHz repetition rate. On the ground, by post-processing the received quantum states at an average of 0.14 photons/pulse, clock data recovery and polarization reference-frame synchronization were successfully done even under remarkable Doppler shifts. A quantum bit error rate below 5% was measured, demonstrating the feasibility of quantum communication in a real scenario from space.**

Recently, the increasing availability of small-size satellites as well as low-cost launches is leading to a rapid growth in the number of satellite-constellation programs, in which thousands of satellites orbiting in a Low Earth Orbit (LEO) work in concert with each other for remote sensing and communications with coordinated ground coverage. Data-intensive satellite sensors mounted in such a constellation produce a large amount of information to be transmitted to the ground in a short time, which requires high capacity communications. However, conventional satellite communications based on microwave frequency bands will struggle to provide the needed capacity because these bands are already congested and severely regulated, and hence the frequency licensing process is lengthy. In the last decade, laser communication (lasercom) has evolved as a promising alternative for high-capacity data links from space [1-11], overcoming microwave communication in several key aspects, such as much higher data rates, being able to use an unregulated spectrum, ultra-low inter-channel interference, smaller and lighter terminals, and power-efficient transmission. In fact, the feasibility of satellite lasercom has been demonstrated by many space missions so far. However, they were based on dedicated bulky satellites of large size, typically several hundred kg with the lasercom terminal mass over 10 kg.

Information security is also becoming an urgent issue in satellite constellations, because the amount of critical and valuable data to be communicated is increasing. Space quantum communication can enhance not only the overall performance of conventional lasercom, but also provides a prerequisite platform for intrinsically hack-proof secure communication, i.e., QKD [12]. The satellite QKD technology actually allows a global scale QKD, which cannot be covered only by earthbound networks due to inevitable losses of optical fibers. There have been significant efforts on developing the basic technologies for QKD in space, namely, terrestrial free-space quantum-communications [11,12-14], demonstrations with moving terminals to emulate the motion of a satellite [15,16], experiments using orbiting objects such as passive corner-reflector satellites to receive single photons in a ground station [17,18], a program to miniaturize the QKD technologies for future cube satellite missions [19], and an experiment of quantum-limited coherent communication from a geostationary satellite to a ground station [20]. Recently a 600-kg quantum-communication satellite has been launched into an orbit for QKD and quantum teleportation experiments [21]. However, it remains a greater challenge to demonstrate quantum communication with a small-size and low-cost satellite. If this could be done using a micro-satellite, the paradigm of satellite communications would change.

Here, we report the first satellite-to-ground quantum-communication experiment using a micro-satellite based on polarization encoding and the ground station with single-photon counters. The SOTA (Small Optical TrAnsponder) terminal, which is as light as 5.9 kg, onboard the micro-satellite SOCRATES (Space Optical Communications Research Advanced Technology Satellite), which is as small as 50 cm cube and 48 kg, transmitted Pseudo-Random Binary Sequences (PRBSs) of non-orthogonal polarization states at a 10-MHz repetition rate from a LEO using a wavelength of 0.8 μm. On the ground, the polarized quantum states at an average of 0.14 photons/pulse were received by a 1-m diameter telescope. By post-processing the received sequence of quantum states, clock data were successfully recovered even under remarkable Doppler shifts, and the polarization reference frame could be well aligned between SOTA and the ground station. Binary non-orthogonal polarization states were finally discriminated by a polarizing quantum receiver with the Quantum Bit Error Rate (QBER) below 5%, demonstrating the feasibility of quantum communication (e.g. B92 QKD protocol [22]) in a real scenario from space.

The SOTA lasercom terminal was designed to carry out feasibility studies on optical downlinks and quantum communications with a low-cost platform onboard the micro-satellite SOCRATES inserted in a LEO at an altitude of about 650 km. Instead of a fine-pointing mechanism, which usually requires an additional bulky payload, the transmission of the 0.8-μm signals is based solely on a coarse-tracking gimbal system with stepping motors. So far, optical downlinks of imaging-sensor data at different wavelengths (980 nm and 1550 nm) were successfully carried out, using On-Off Keying (OOK) modulation at 10 Mbit/s from SOTA, and a 1-m diameter telescope in the Optical Ground Station (OGS) at the NICT headquarters, in Koganei (Tokyo, Japan). An experiment on the effect of the atmospheric propagation on the polarization was performed using circular and linear polarizations transmitted from SOTA [11]. CNES (National Centre for Space Studies) could also successfully receive the signals from SOTA in the 1.54-m MeO OGS in Caussol (France), demonstrating satellite-to-ground links with adaptive optics to compensate atmospheric effects [9,10]. These works demonstrated the technological maturity of optical acquisition, tracking and pointing, and clock synchronization in the classical optical regime, based on a 50-kg-class micro-satellite.

After the success of these experiments, we moved forward to the quantum-communication experiment based on binary non-orthogonal linear-polarization encoding to emulate the B92 QKD protocol [22]. The purpose of the quantum-communication experiment was to verify the feasibility of the polarization-encoded onboard-laser-transmitter technology in orbit and the photon-counting polarization-decoding technology on the ground through a space-to-ground slant atmospheric path. Since the laser beam divergence was widened to be able to track the OGS more reliably with the SOTA coarse pointing, brighter laser pulses than those required in QKD were used, although the optical signals arriving at the NICT OGS were photon-limited at 0.14 photons/pulse on average, which is in the regime of quantum communication.

The polarization encoding is the most reasonable option for quantum communications from space thanks to its stable propagation through the atmosphere while time-bin encoding is widely used in fiber networks [23]. The polarization-based quantum communication allows simple and compact implementations of transmitter and receiver systems with low-cost optical components because it does not require an interferometer, and hence can be adapted to an environment of strong mechanical vibrations. A big challenge in this kind of systems is the polarization reference-frame synchronization between the fast-moving LEO satellite and the OGS in order to perform a reliable implementation of the QKD protocol. Another challenge is the clock-data recovery using the received sequences of quantum states directly, which will enable compact and low-cost transmitter and receiver implementations. For a slant-atmospheric downlink from a LEO satellite, the Doppler shift is also an important factor to evaluate for precise clock-data recovery. The purposes of this work are to solve these two main issues in order to demonstrate correct decoding of polarized quantum-state sequences and finally be able to evaluate the QBER.

For these purposes, we transmitted repeating PRBSs generated by a linear feedback shift register with a period of $2^{15} - 1 = 32767$, the so called PN15, encoding it into a signal sequence of binary non-orthogonal polarization states. Using this known bit pattern of PRBSs, we performed the necessary tasks for quantum communication, including clock data recovery, timing offset identification, bit pattern synchronization, polarization reference-frame synchronization, and decoding of the polarized quantum states.

Figure 1 shows a picture of SOTA (Fig. 1a), the configuration of the two linearly polarized laser diodes Tx2 and Tx3 (Fig. 1b) in SOTA, the receiver telescope and the quantum receiver in the NICT OGS (Fig. 1c). The linear polarizations of Tx2 and Tx3 are aligned at a difference of -45° (the actual separation angle is -44°). Tx2 and Tx3 are driven by the PN15 PRBSs, based upon a clock frequency of $f_0 = 10$ MHz, with a Unit Time Interval (UTI) of $T_0 = 100$ ns. Tx2 emits a horizontally polarized pulse (H) synchronized with every rising edge of the PRBS, and Tx3 emits a -45° polarized pulse (-45°) synchronized with every falling edge (see the upper part of Fig. 3c shown later on). Thus, the transmitted signal consists of sequences of H- and -45°-polarization states.

**Transmitter and receiver**

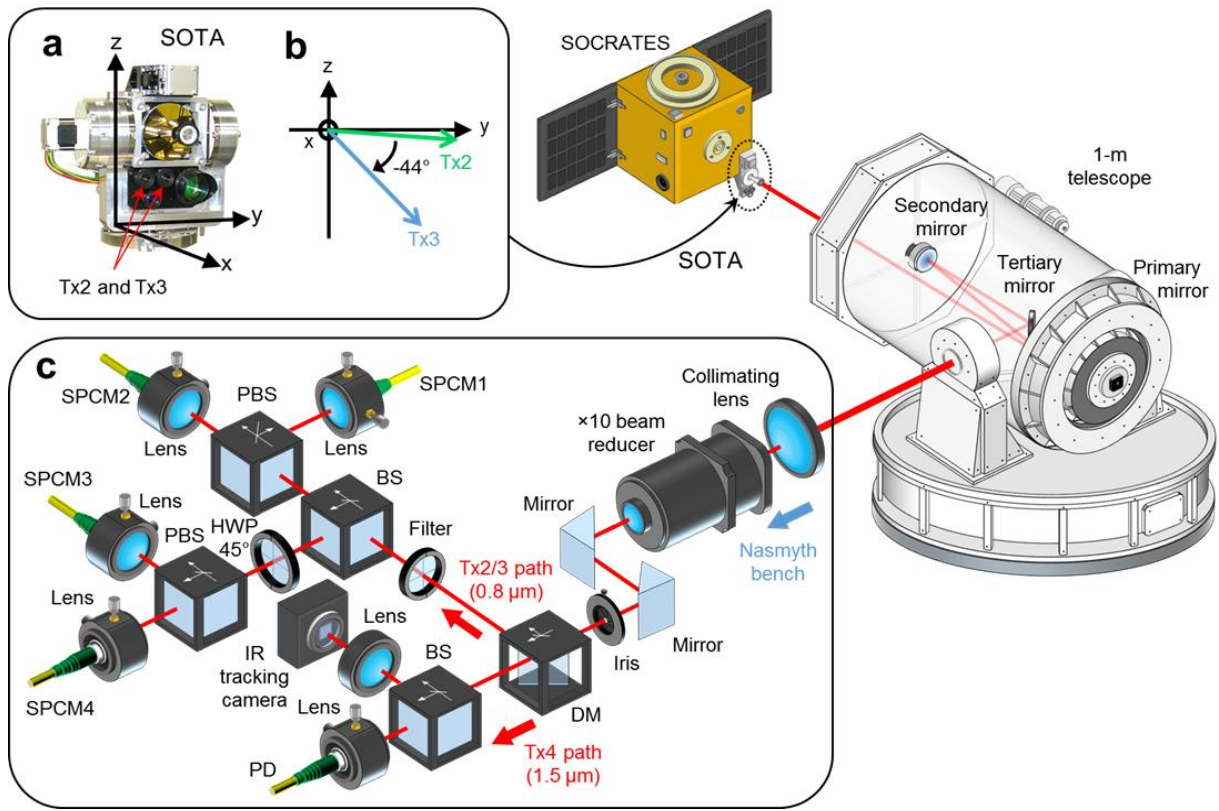

**Figure 1. Transmitter and receiver systems.** (**a**) A picture of SOTA, whose size is 18 cm width x 11 cm depth x 27 cm height. (**b**) Configuration of the two linearly polarized laser diodes Tx2 and Tx3 in SOTA. (**c**) The receiver telescope and the quantum receiver in the NICT OGS. DM; dichroic mirror; PD, photodetector; IR, Infrared; BS, beam splitter; PBS, polarizing beam splitter; HWP, half wave plate; SPCM, single-photon counter module.

The NICT OGS consists of the 1-m diameter Cassegrain telescope and the polarization-based quantum receiver. The incident light reflected by the primary and the secondary mirror passes through a tertiary mirror, made of aluminum to minimize the linear polarization deterioration. At this point, the beam after the tertiary mirror has a width of 3 mm, and is guided towards the quantum receiver installed at the Nasmyth bench of the telescope. A 1.5-μm-wavelength circularly-polarized laser beam transmitted from SOTA was used for satellite tracking purposes, and was separated from the 0.8-μm light using a dichroic mirror in the quantum receiver. This 1.5-μm beam was then guided to a photodetector and monitored using an IR camera.

The quantum receiver consists of beam splitters, polarizing beam splitters and half wave plates, ending with four ports, where four Single-Photon Counter Modules (SPCMs of Excelitas Technologies Corp.) were used as detectors after coupling the beams to multi-mode optical fibers using converging lenses. Received photon counts were then time-tagged by a time-interval analyzer (Hydraharp 400 of PicoQuant) whose timing resolution is 1 ps, generating a time-tagged photon-count sequence per each SPCM. Table 1 shows the losses of the quantum receiver, the estimated atmospheric attenuation and total loss budget for a 50° elevation angle. The total loss includes the coupling loss of the arriving beam into the receiver telescope and the following losses: the quantum receiver loss including the SPCM coupling efficiency as the main source of loss; the receiver's telescope loss including the secondary and tertiary mirrors reflectivity and a vignetting effect produced in the secondary mirror because it was replaced by a slightly smaller uncoated aluminum mirror to maintain the linearity of the Tx2/Tx3 polarizations. The atmospheric attenuation was calculated with MODTRAN (MODerate resolution atmospheric TRANsmission by Spectral Sciences) using the conditions of the quantum experiment.

Table 1. Receiver losses for a 50° elevation angle.

|  | Tx2/Tx3 |
| --- | --- |
| **Quantum receiver loss** | -16 dB |
| **Receiver's telescope loss** | -1 dB |
| **Secondary/tertiary mirrors loss** | -1.68 dB |
| **Atmospheric attenuation** | -3 ~ -6 dB |
| **Total loss budget including the above and coupling losses** | -78.18±8.5dB |

**Clock-data recovery and timing-offset identification**

In the OGS, the clock data was first recovered by post-processing a part of the received photon-count sequence from the quantum receiver at the OGS. Due to the heavy attenuation in the atmospheric path of optical downlink and in the quantum receiver, which is roughly -87dB~-70dB in total, many optical pulses emitted at SOTA did not arrive at the SPCMs in the quantum receiver. Therefore a 10-Mbit block of the transmitted PRBS at SOTA was used for the clock-data recovery. Since SOCRATES was moving fast at a velocity of 7 km/s with propagation distances ranging from 650 km to 1000 km, the Doppler shift through each optical link campaign was expected to be within ±200 Hz around the clock frequency $f_0$ = 10 MHz with the shifting rate (frequency drift) of 3 Hz/s. Thus the UTI and the clock frequency at the OGS deviate from those at SOTA. This effect should be taken into account in the clock-data recovery. To this end, we analyzed a received photon-count sequence for 1sec (a time to acquire a 10-Mbit block of PRBS), from 22:59:00 to 22:59:01 JST on 5[th] of August 2016, by using various possible clock frequencies (equivalently UTIs) to find the values which best fit the received photon-count sequence. The total counts in this sequence were 7119.

The clock frequency and the frequency drift at the OGS which best fitted the received photon-count sequence were identified to be $f$ = 10,000,096.5 Hz and -2.25 Hz/s, respectively. The timing offset of the received photon-count sequence from the transmitted PRBS could also be found as 0.64 with the full width of ±0.03 in the normalized UTI. This time window of 0.61~0.68 UTI can then be used for selecting the signal counts out of the noise counts. By this time-gating, 816 counts could be subtracted as dark counts from the total counts of 7119. As a result, the effective dark count rate was reduced to about 10c/s per photon counting channel, which is roughly 10% of the intrinsic SPCM dark count rate.

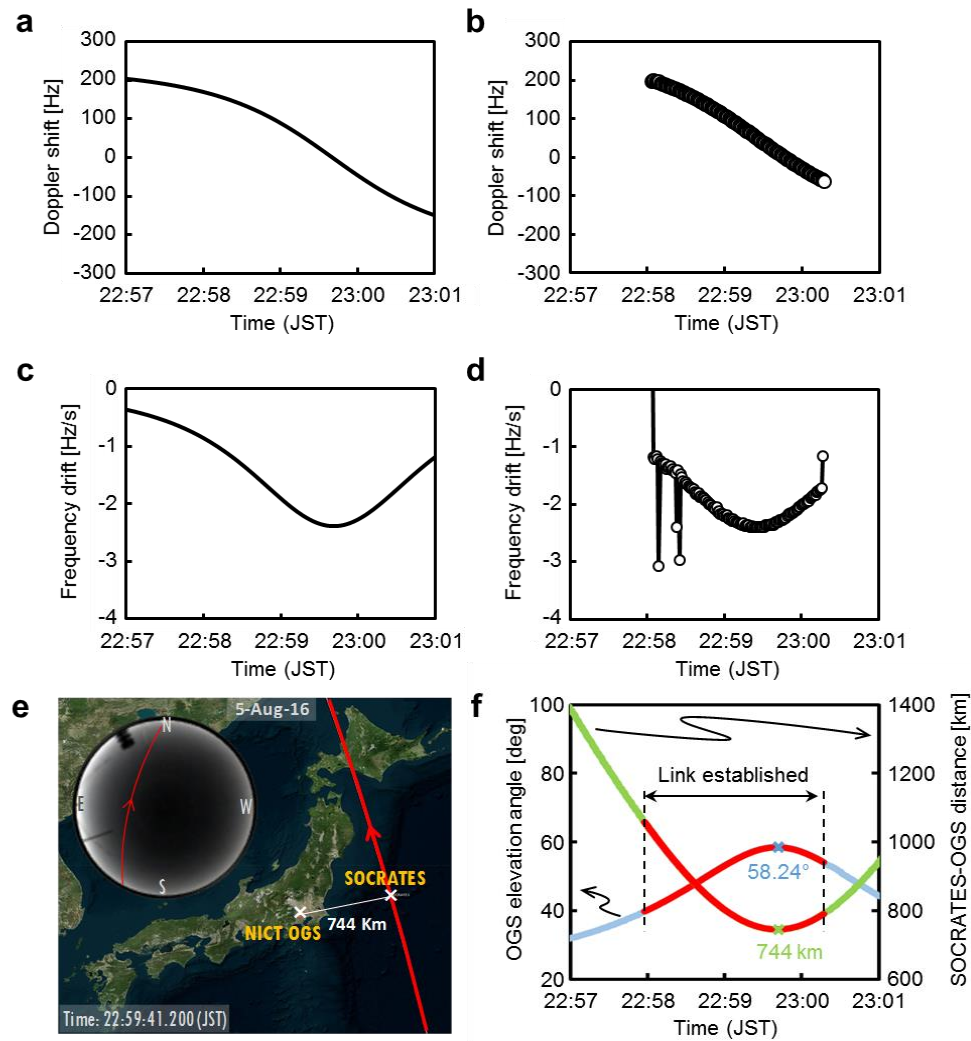

**Figure 2. The Doppler shifts and the frequency drifts in the optical link campaign on 5$^{th}$ of August 2016.** (**a**) Predicted Doppler shift. (**b**) Observed Doppler shift. (**c**) Predicted frequency drift. (**d**) Observed frequency drift. (**e**) Configuration of the SOCRATES orbit and the NICT OGS. (**f**) Elevation angle from the OGS telescope to SOCRATES as well as the distance between SOCRATES and the OGS.

Figure 2 represents the Doppler shift (Figs. 2a and 2b) and the frequency drift (Figs. 2c and 2d) evaluated for the optical-link campaign for a 2-min-15-sec duration from 22:58:03 to 23:00:18 on 5$^{th}$ of August 2016. Figures 2a and 2c show the calculated curves from the SOTA orbit on that date, while Figs. 2b and 2d show the observed data. The observed Doppler shift was less than 200 Hz. Considering the possibility that the SOTA clock frequency could also be slightly different from 10 MHz, the time when the frequency drift reached its minimum value is considered to be the time when SOCRATES was at the closest distance from the OGS. These results demonstrate that the clock-data recovery and the timing-offset synchronization could be successfully carried out by using data extracted directly from the received quantum states.

**Bit-pattern synchronization**

After the clock-data recovery and the timing-offset identification, the time-tagged photon-count sequence in the time domain turns into a simple bit sequence in the bit domain. The next task is to establish the synchronization of bit patterns between the transmitted signal sequence and the received bit sequence. This could be made by calculating the cross correlation between the transmitted signal sequence and the received bit sequence for the period of PN15 PRBS, namely the 32767 bit length. Figures 3a and 3b present the experimental result on the cross correlation between the transmitted signal sequence and the received bit sequence. This has a peak at 29656, which corresponds to the offset.

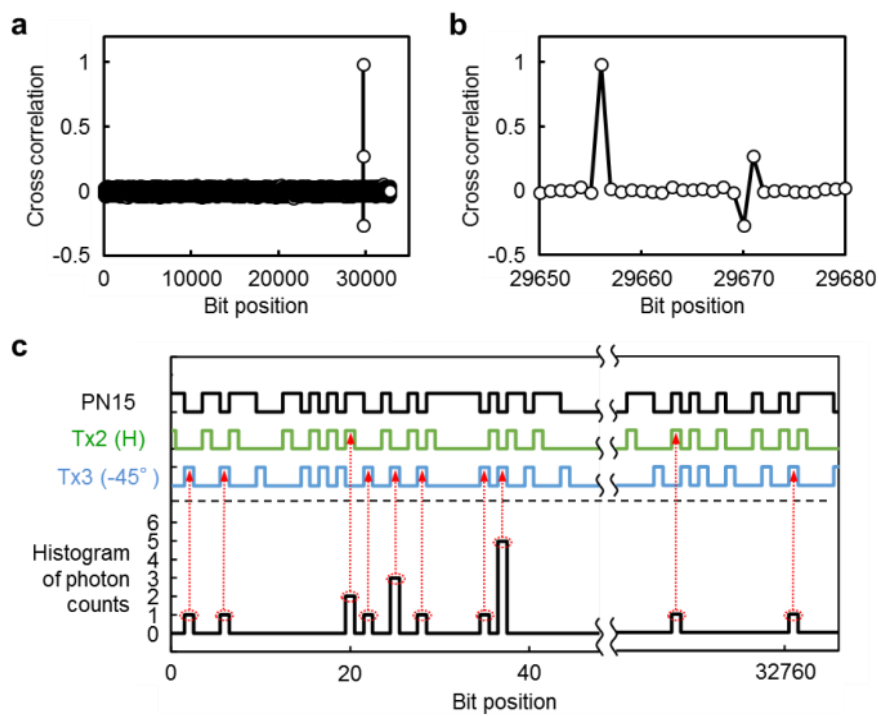

**Figure 3. Experimental result on bit pattern synchronization.** (**a**) The cross correlation between the transmitted signal sequence and the received bit sequence for the period of PN15 PRBS, i.e., the 32767 bit length. (**b**) A magnified plot of the cross correlation around the offset. (**c**) The PN15 PRBS (top), the on/off sequences of Tx2 (second top), Tx3 (third top), and the histogram of photon counts summed up for 1 sec (a time span of 10 Mbit transmitted sequence). Tx2 emits a horizontally-polarized pulse synchronized with every rising edge of the PRBS, and Tx3 emits a -45°-polarized pulse synchronized with every falling edge.

By compensating this offset, we could finally synchronize the photon-count histogram in the bit domain with the transmitted bit patterns. Figure 3c shows, in the bottom, the histogram of photon counts summed up for 1 sec (a time span of a 10 Mbit transmitted sequence) after the bit patterns were synchronized, as well as the PN15 PRBS (top), the on/off sequences of Tx2 (second top), and Tx3 (third top). As indicated by red vertical arrows, one can recognize that whenever the quantum receiver registered finite counts, SOTA had always emitted optical pulses.

**Polarization reference-frame synchronization**

The tracking of the polarization between SOTA and the OGS should be able to be realized by adaptively rotating a half-wave plate at the entrance of the quantum receiver, based on the orbital and telemetry information from the satellite. However, in the campaign on 5$^{th}$ of August 2016, this polarization-tracking system was not used, and shortly after that the satellite operation of SOCRATES was terminated. In addition, the alignment of the receiver was not optimum because the quantum-communication experiment was carried out in parallel with other experiments within the SOTA mission. Therefore, we performed the polarization reference-frame synchronization by post-processing a part of photon counts registered at the SPCMs.

In principle, this task, whose purpose is to extract the polarization angle of the input state in front of the receiver telescope, could be done with two SPCMs with a polarizing beam splitter. However, because of a non-optimum alignment in this experiment, the situation was more complex. In fact, the polarization characteristics and hence the effective coupling efficiency of the input state into each port to the SPCM in the quantum receiver depends on the telescope elevation and azimuth angle. Therefore we first made a calibration chart to correct the relative sensitivity of each port for various telescope elevation and azimuth angles, by observing the light from different high-luminosity stars as reference point sources. As the light from the stars is not polarized, a rotating linear polarizer was inserted between the telescope and the quantum receiver to prepare linearly polarized light at different azimuth/elevation angles. Then given a known sequence of polarized states either from Tx2 or Tx3, we analyzed photon-count sequences from SPCM1, SPCM2, and SPCM4, because this combination maximized the signal-to-noise ratios in the polarization angle evaluation, and finally reconstructed the polarization angle which varied as SOTA was moving. This evaluation was performed for every second during the optical-link campaign from 22:58:03 to 23:00:18 on 5$^{th}$ of August 2016.

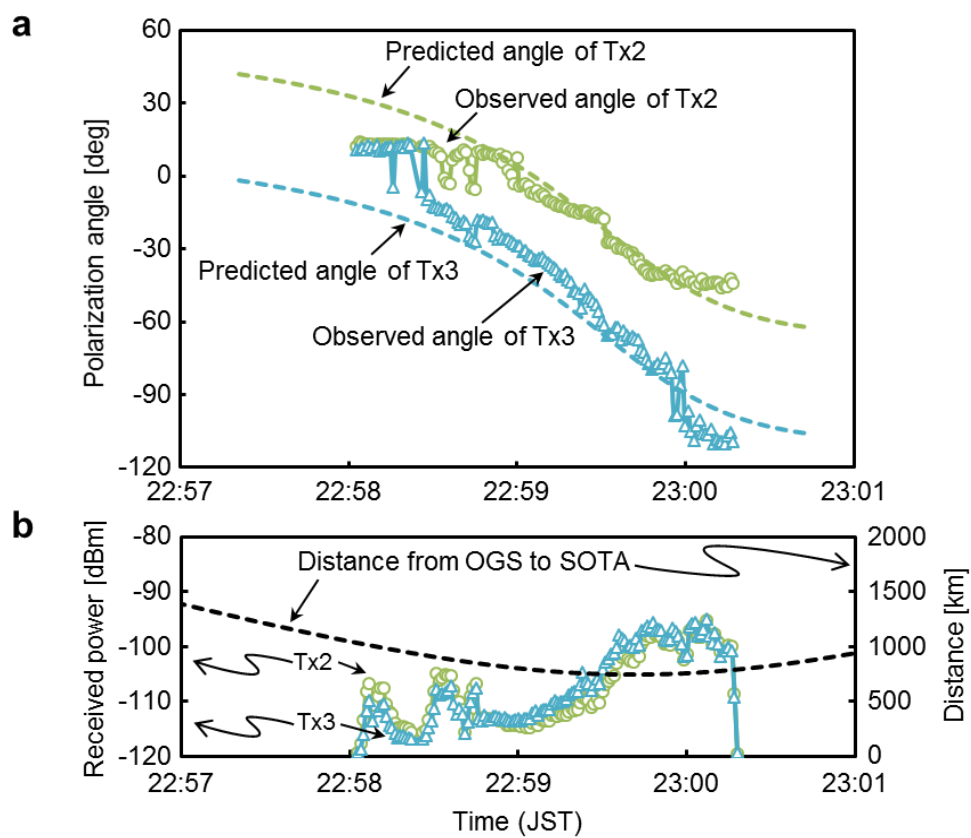

**Figure 4. Experimental results on polarization angle variations and received power measured for Tx2 and Tx3 sequences.** (**a**) The reconstructed variations of polarization angle in front of the receiver telescope for Tx2 and Tx3 (solid curves), and theoretical curves calculated by the orbital information of SOCRATES (dashed curves). (**b**) The variations of received power estimated by the total photon counts of the SPCMs for the sequences from Tx2 and Tx3, as well as the distance from the NICT OGS to SOTA.

Figure 4a shows, by solid curves, the reconstructed variations of the two polarization states in front of the receiver telescope. The two curves for Tx2 and Tx3 were derived independently by the above procedure of polarization reference-frame synchronization. The dashed curves are theoretical curves calculated by the orbital information of SOCRATES. Figure 4b shows the variations of received power estimated by the total photon counts of the SPCMs for the sequences from Tx2 and Tx3, as well as the distance from the NICT OGS to SOTA. In the first half period 22:58:03~22:59:00, the received power fluctuated, which may be due to unstable tracking. On the other hand, in the last half period 22:59:00~23:00:18, the tracking could be more stabilized, the received power was increasing, and the observed curves of polarization angle variation could be fitted much better with the theoretical curves.

**Quantum bit error rate**

Once the polarization reference-frame was established between SOTA and the OGS, we can estimate an essential parameter for the QKD protocol, i.e., QBER. In the B92 protocol which we emulate here, the bit information 0 and 1, denoted as inputs $x$=0, 1, are encoded into binary non-orthogonal quantum states. According to quantum mechanics, it is impossible to distinguish them with certainty. Moreover, the more one tries to distinguish the two states, the more the states get disturbed. These quantum states are detected by a receiver which has the three kinds of outcomes; (i) the detection of the 0 with certainty, (ii) the detection of the 1 with certainty, and (iii) the inconclusive outcome for which both possibilities of the 0 and the 1 are implied, which is dealt with as failure events. These outcomes are denoted as $y$=0, 1, and F. The inconclusive outcomes, $y$=F, are discarded, whose process is referred to as the sifting. After the sifting, one can get the transition statistics, $N(y|x)$, which represents the number of events detecting $y$ given the input $x$.

The QBER is given by

$$QBER = \frac{N(1|0)+N(0|1)}{\sum_{y=0,1} N(y)}$$

where $N(y) = \sum_{x=0,1} N(y|x)$.

Since we did not track the polarization angle of SOTA but established the polarization reference-frame by post-processing in our experiment, we chose, for the QBER estimation, a 12-sec duration of 22:59:21~22:59:33 on 5$^{th}$ August 2016, in which the quantum states arrived at the OGS were actually -45° and -90°, which were originally emitted in H from Tx2 and in -45° from Tx3 at SOTA, respectively. For this configuration, the three outcomes in the quantum receiver correspond to the clicks at (i) SPCM3 for $y$=0, (ii) SPCM2 for $y$=1, and (iii) SPCM1 or SPCM4 for $y$=F (see Fig. 1 again for SPCM positions). The observed QBER was smaller than 4.9% and reached the minimum value of 3.7% at 22:59:25. The variation of QBER around this time is shown in Fig. 5 for a wider duration of 1 min (22:59:00~23:00:00). The increase of QBER outside the estimation interval as marked by vertical dashed lines are attributed to the non-optimal polarization reference-frame configuration of the quantum receiver for the input polarized quantum states, which was caused because we could not use the polarization tracking between SOTA and the OGS.

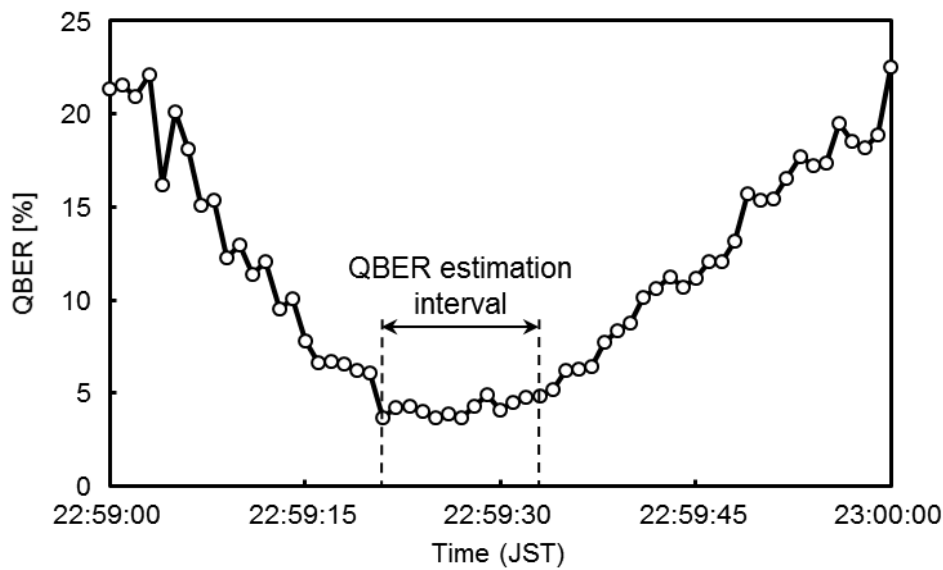

**Figure 5. The variation of QBER in the emulated B92 protocol for a 1-min duration of 22:59:00 ~ 23:00:00 on 5th August 2016.**

**Conclusion**

We have demonstrated the first satellite-to-ground quantum-communication experiment with a micro-satellite as small as 50-kg-class. Our techniques of clock-data recovery, and polarization reference-frame synchronization directly from quantum states will enable compact implementation of a quantum communication system. The results of QBER<5% demonstrated the feasibility of quantum communication in a real scenario from space.

**Acknowledgements**

The authors acknowledge Maki Akioka, Toshihiro Kubooka, and Hiroyuki Endo for their technical supports on SOCRATES operation and data analysis. This work was supported in part by the ImPACT Program of the Cabinet Office Japan.